\documentclass[1p]{elsarticle}
\usepackage{amsmath}
\usepackage{epsfig}
\usepackage{subfigure}
\usepackage{float}
\usepackage{longtable}
\usepackage{times}
\usepackage{txfonts}
\usepackage{color}
\usepackage{ulem}

 \usepackage{ifpdf}
 \ifpdf
 \usepackage[pdftex]{hyperref}
 \else
 \usepackage[hypertex]{hyperref}
 \fi

 \hypersetup{
   pdftitle={},%
   pdfauthor={},%
   pdfsubject={},%
   pdfkeywords={},%
   pdfstartview={},%
   bookmarksopen=true, breaklinks=true, debug=true, %
   colorlinks=true, linkcolor=red, citecolor=blue, urlcolor=blue
 }

\journal{Nuclear Physics A}

\newcommand{\beq}{\begin{eqnarray}}
\newcommand{\eeq}{\end{eqnarray}}
\newcommand{\be}{\begin{eqnarray*}}
\newcommand{\ee}{\end{eqnarray*}}

\newcommand{\eqs}[1]{\begin{equation} \begin{split} #1\end{split} \end{equation} }

\newcommand{\eg}{{\it e.g.}}

\newcommand{\ce}[1]{Eq.~\eqref{#1}}

\newcommand{\cf}[1]{{Fig.~\ref{#1}}}

\def\lsim{\raise0.3ex\hbox{$<$\kern-0.75em\raise-1.1ex\hbox{$\sim$}}}
\def\gsim{\raise0.3ex\hbox{$>$\kern-0.75em\raise-1.1ex\hbox{$\sim$}}}

\def\pPb  {$p$Pb}

\def\Ncoll   {\mbox{$N_{\rm coll}$}}

\def\beq     {\begin{equation}}
\def\eeq     {\end{equation}}

\long\def\symbolfootnote[#1]#2{\begingroup%
  \def\thefootnote{\fnsymbol{footnote}}\footnote[#1]{#2}\endgroup}

\begin{document}

\begin{frontmatter}

\title{Open-beauty production in $p$Pb collisions at 
$\sqrt{s_{NN}}=5$ TeV: effect of the gluon nuclear densities}

\author
{Z. Conesa del Valle$^a$, E. G. Ferreiro$^b$, F. Fleuret$^c$, J.P. Lansberg$^a$, A. Rakotozafindrabe$^d$}

\address{
$^a$IPNO, Universit\'e Paris-Sud, CNRS/IN2P3, F-91406, Orsay, France \\
$^b$Departamento de F{\'\i}sica de Part{\'\i}culas and IGFAE, Universidade de Santiago de Compostela, 
15782 Santiago de Compostela, Spain\\
$^c$Laboratoire Leprince Ringuet, \'Ecole polytechnique, CNRS/IN2P3,  91128 Palaiseau, France\\
$^d$IRFU/SPhN, CEA Saclay, 91191 Gif-sur-Yvette Cedex, France}

\begin{abstract}
We present our results on open beauty 
production in proton-nucleus collisions for the recent
LHC \pPb\ run at  $\sqrt{s_{NN}}=5$ TeV.  
We have analysed the effect of the modification of the gluon PDFs in nucleus 
at the level of the nuclear modification factor. 
Because of the absence of measurement in $pp$ collisions at the same energy $\sqrt{s_{NN}}$,
we also propose the study of the forward-to-backward yield ratio in which the unknown proton-proton yield cancel.
Our results are compared with the data obtained by LHCb collaboration and show a good agreement.
\end{abstract}


\end{frontmatter}


\section{Introduction}
\label{sec:intro}

The recent $p$Pb run at 5 TeV which has taken place at the CERN LHC provides exciting measurements that need to be
interpreted from the phenomenological point of view.
In particular, the LHCb collaboration has released the first result of the nuclear matter effects on
beauty production in $p$Pb collisions at 5 TeV at the LHC through the study of non-prompt $J/\psi$. Since the data cover
the low $p_T$ region down to zero, they really constitute a direct access to open beauty production in $p$Pb
collisions\footnote{It is important to keep in mind that the $J/\psi$'s are produced by weak decay way outside the
nucleus.}.

\vskip 0.15cm
In these proceedings, we follow the lines of
our previous studies devoted to the investigation of the cold nuclear matter effects,
as the nuclear attenuation and the modification of the gluon densities,
on 
quarkonium production in proton-nucleus collisions, 
both at RHIC \cite{Ferreiro:2008wc,Ferreiro:2009ur,Ferreiro:2012mm,Ferreiro:2012sy,Rakotozafindrabe:2012ss} and LHC \cite{Ferreiro:2011xy,Ferreiro:2013pua} energies.

\vskip 0.15cm
In the following, we present our results on the nuclear modification factor and on the forward-to-backward ratio for open-beauty production
using LO pQCD for $b\bar{b}$ production and the nuclear PDF (nPDF) fit EPS09 and nDSg at LO accurancy.


\section{Theoretical framework}

As for our precedent studies on prompt $J/\psi$ and $\Upsilon$ production in proton-nucleus collisions at RHIC and LHC energies, 
we have used our probabilistic Glauber Monte-Carlo framework,
{\sf JIN}~\cite{Ferreiro:2008qj},
which allows us to encode different mechanisms for the partonic production and to interface these production processes
with different cold nuclear matter effects, such as the shadowing,
in order to get the production cross sections for
$pA$ and $AA$ collisions.

\vskip 0.15cm
As far as the open beauty partonic production is concerned, we have used the LO calculations from 
\cite{Combridge:1978kx} which applies for the production of $b\bar{b}$ quarks in proton-proton collisions.

\vskip 0.25cm
Other approaches, as the successful NLO
prediction of the FONLL \cite{Cacciari:2003uh}
which describes well the beauty
production at Tevatron and LHC energies can obviously also be used. However, 
a comparison of our LO result with more refined theoretical approaches
\cite{desy} 
shows that the LO description is sufficient to describe the low $p_T$ cross section up to $1-2 m_b$.

\vskip 0.15cm
Consequently, we believe that a LO evaluation is sufficient  to determine the  proton-proton production 
kinematics to be used in our Glauber code.
Note that, in order to compare our results on $b$ quarks with the data for $J/\psi$ from $b$, we will 
make the approximation that they occur at the same rapidity. This approximation is reasonable as can 
be seen on \cf{fig:figzaida} which shows the rapidity correlation between the rapidity of the $B$ and 
that of the $J/\psi$ decay product.
\begin{figure}[H]
\begin{center}\vspace*{-0.1cm}
{\includegraphics[trim = 0mm 0mm 0mm 0mm, clip,height=5.5cm]{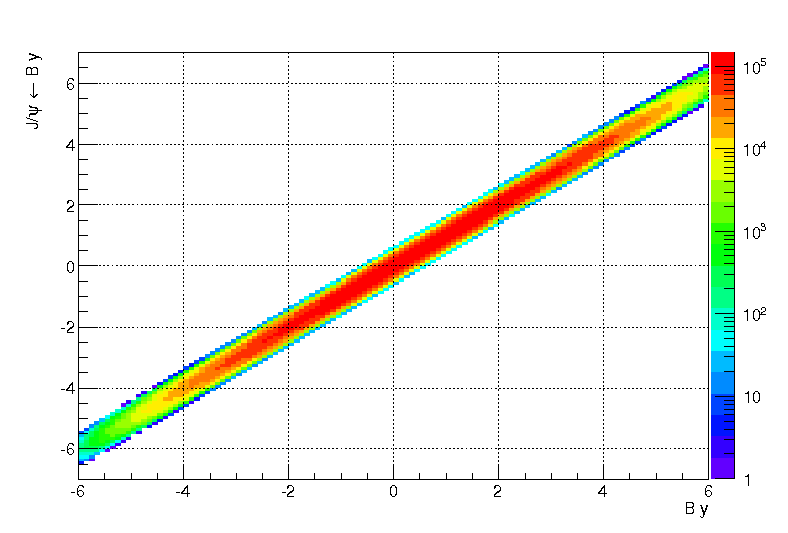}}\\\vspace*{-0.cm}
{\includegraphics[trim = 0mm 0mm 0mm 0mm, clip,height=5.5cm]{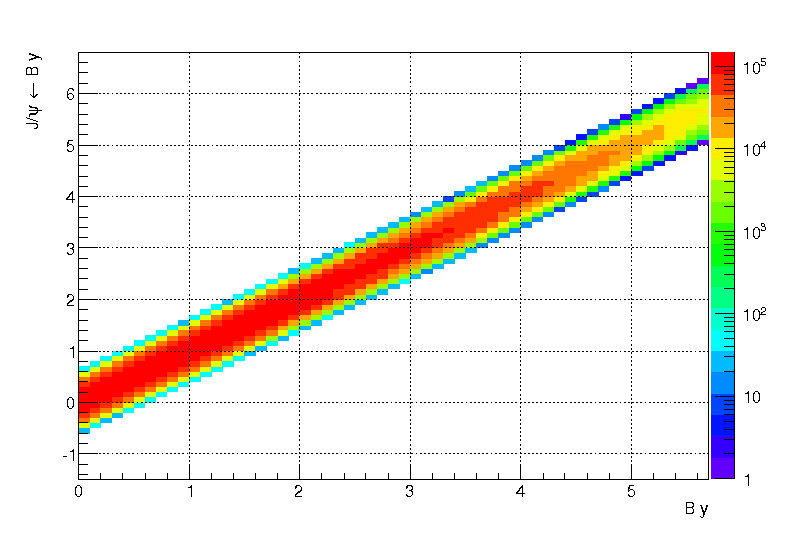}}\vspace*{-0.4cm}
\end{center}
\caption{(Colour online) Rapidity correlation between $B$ hadrons and the $J/\psi$
coming from their decay at 5 TeV.
}\label{fig:figzaida}\vspace*{-0.4cm}
\end{figure}
We note however that the approximation that the production kinematics of the $J/\psi$ be similar to that of the $b$ quark is only justified because one integrates on $p_T$. One would have to be much more careful if we wanted to analyse $p_T$ dependent effects.

\vskip 0.15cm
Our results for $J/\psi$ from $b$ production in proton-proton collisions, integrated over $p_T$ and 
compared to the experimental data from LHCb Collaboration 
at 7 \cite{Aaij:2011jh} and 8 TeV \cite{Aaij:2013yaa}, are shown in \cf{fig:fig0a}. We have 
introduced the uncertainty related to the choice of the factorisation scale, $\mu_F$, also 
referred to as $Q$, that we have taken to be $(0.75, 1, 2) \times m_T$. 
We obtain a good agreement, as expected.

\begin{figure}[H]
\begin{center}\vspace*{-0.1cm}
{\includegraphics[trim = 0mm 0mm 0mm 0mm, clip,height=4.5cm]{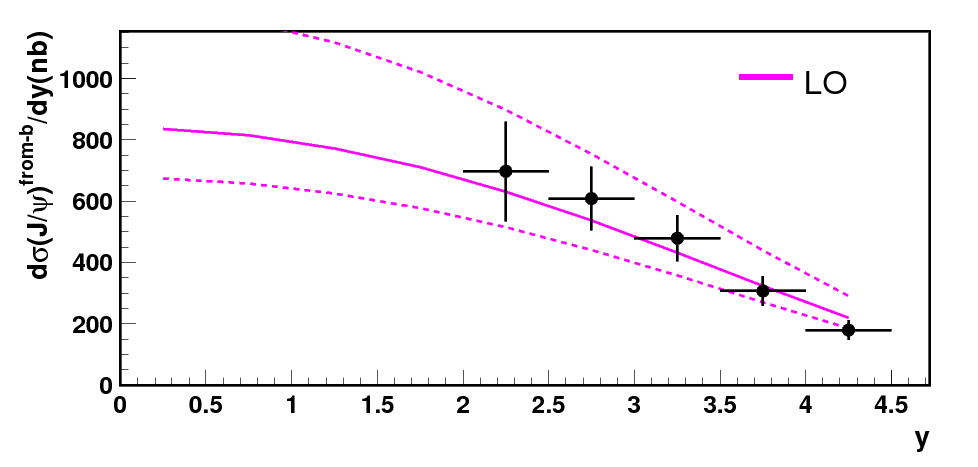}}\\\vspace*{-0.cm}
{\includegraphics[trim = 0mm 0mm 0mm 0mm, clip,height=4.5cm]{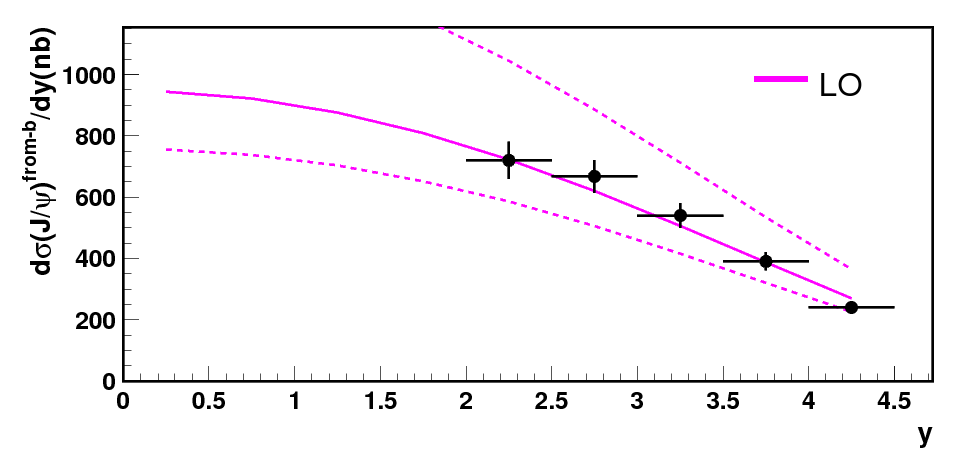}}\vspace*{-0.4cm}
\end{center}
\caption{(Colour online) 
Differential production cross-section as a function of $y$ integrated over $p_T$, for $J/\psi$ from $b$ at 7 TeV (up)
and 8 TeV (down)
compared to experimental data \cite{Aaij:2011jh,Aaij:2013yaa}.
The effect of the unknown factorisation scale
is taken to be $(0.75, 1, 2) \times m_T$ (lower, central and upper curves).
}\label{fig:fig0a}\vspace*{-0.4cm}
\end{figure}

\vskip 0.25cm
In which concerns the nuclear modification of the gluon PDFs, we have employed the parametrisations
EPS09~\cite{Eskola:2009uj} and nDSg~\cite{deFlorian:2003qf} at LO accuraracy.
The spatial
dependence of the nPDF has been taken into account in our approach, assuming an inhomogeneous shadowing
proportional to the local density~\cite{Klein:2003dj,Vogt:2004dh}.
For the results presented here, this $b$-dependence does however not enter.

\vskip 0.15cm
As in \cite{Ferreiro:2013pua}, we use the central curve of EPS09 as well as 
four specific extremed curves (minimal/maximal shadowing, minimal/maximal EMC effect) 
which reproduce the enveloppe of the gluon nPDF uncertainty encoded in EPS09 LO.

\section{Results}
As for any hard probes, the suppression of the open beauty can be characterised 
by the {\it nuclear modification factor}, $R_{pA}$
--the ratio of the yield in $pA$ collisions to the yield in
$pp$ collisions at the same energy multiplied by the average
number of binary collisions in a typical proton-nucleus collision, $\langle \Ncoll\rangle$:
\beq
R_{pA}=\frac{dN_{pA}^{}}{\langle\Ncoll\rangle dN_{pp}^{}}.
\eeq
If a nuclear effect is at work, it leads to a deviation of $R_{pA}$ from {\it unity}.
\begin{figure}[H]
\begin{center}\vspace*{-0.1cm}
{\includegraphics[trim = 0mm 0mm 0mm 0mm, clip,height=6cm]{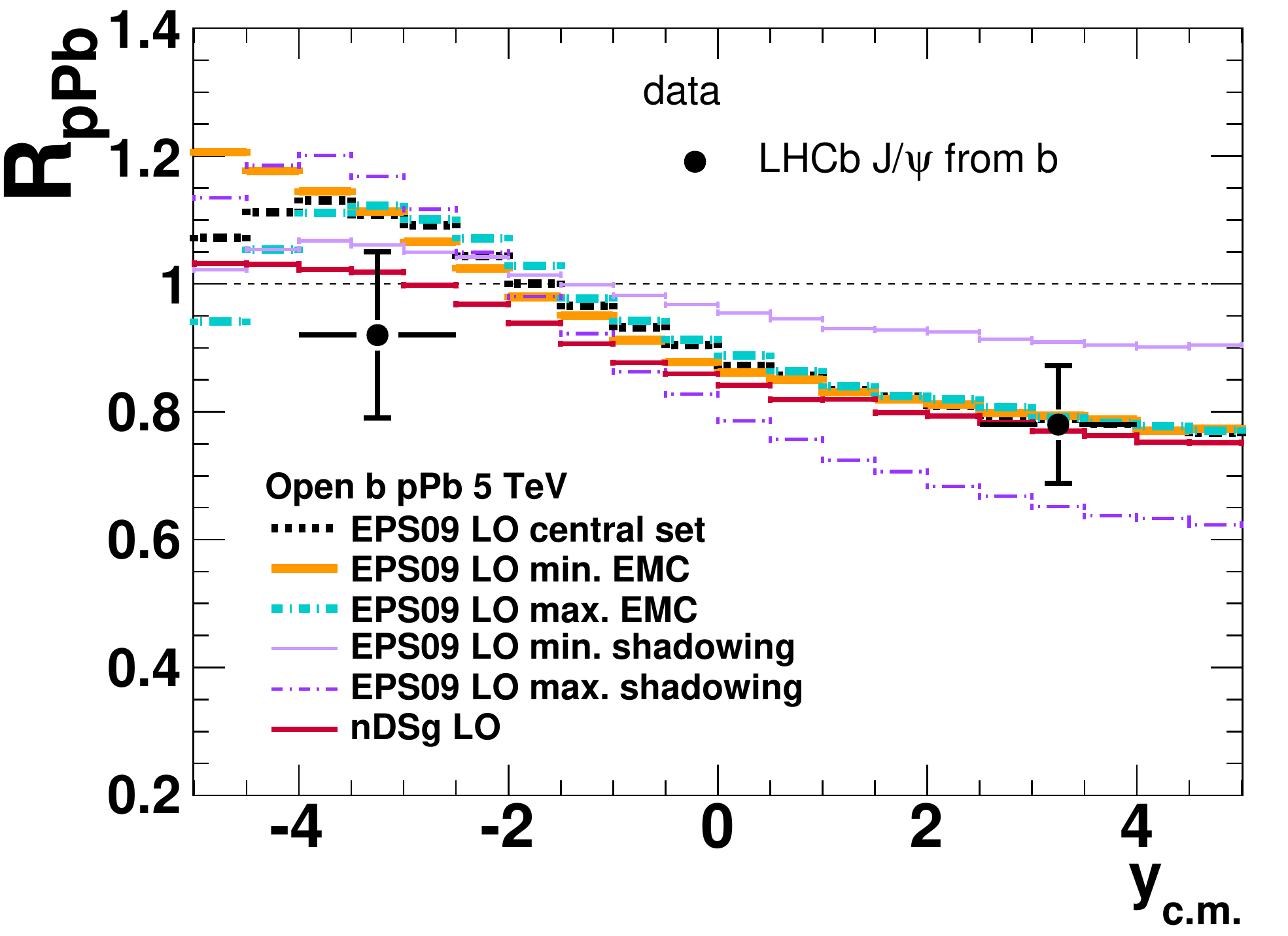}\label{fig:rpPb-ptcut}}\\\vspace*{-0.cm}
{\includegraphics[trim = 0mm 0mm 0mm 0mm, clip,height=6cm]{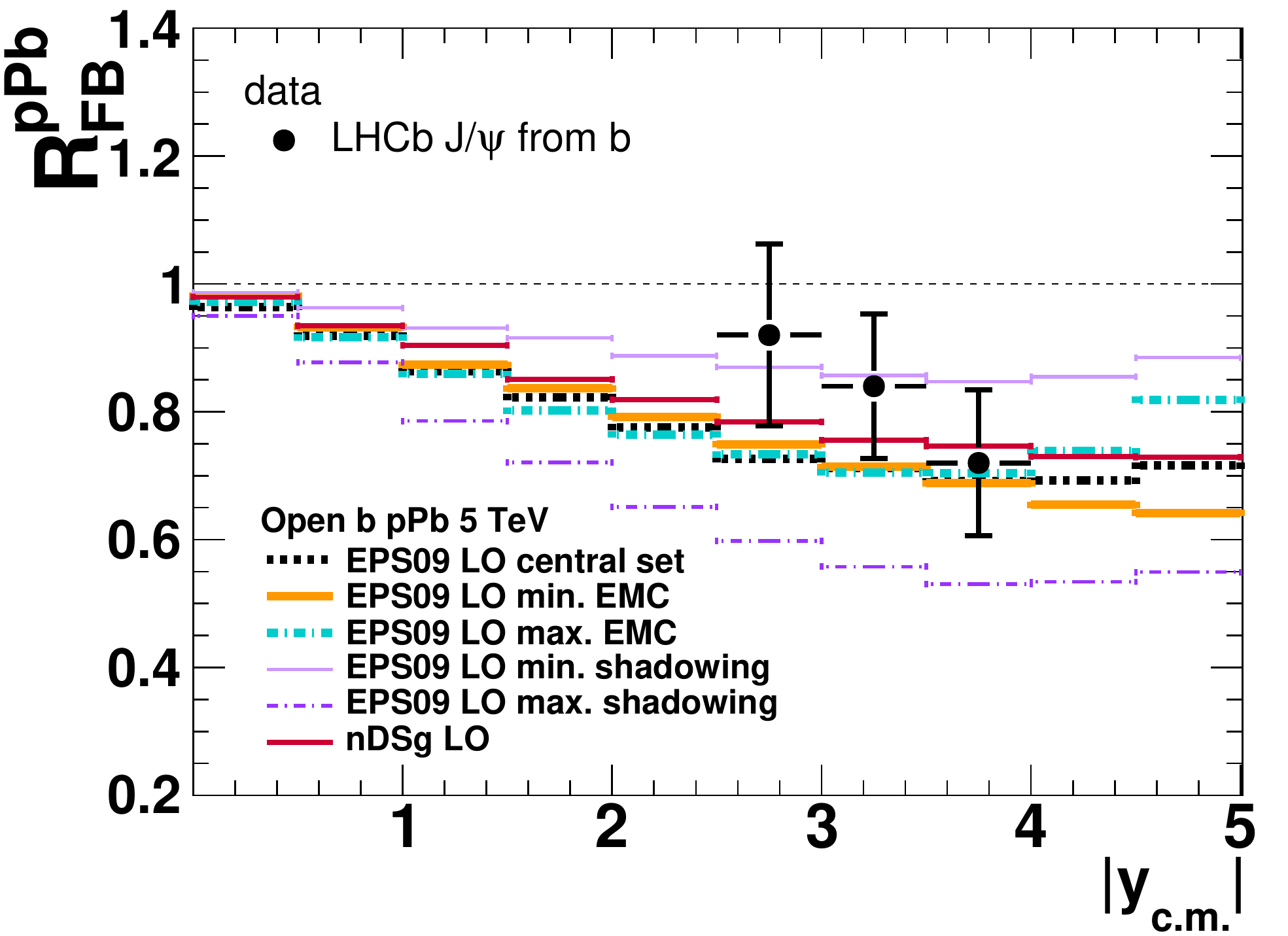}\label{fig:RFB-ptcut}}\vspace*{-0.4cm}
\end{center}
\caption{(Colour online) Open-b nuclear modification factor (up) and its forward-to-backward (down) ratio  in \pPb\ collisions, 
 at $\sqrt{s_{NN}}=5\mathrm{~TeV}$ versus $y$ using 5 extremal curves of the EPS09LO nPDF set and nDSg and compared to experimental results extracted from
LHCb Collaboration\cite{LHCb-CONF-2013-008}.
}\label{fig:fig1}\vspace*{-0.4cm}
\end{figure}

\vskip 0.15cm
It is nevertheless fundamental to insist on the fact that, in the absence of a yield measurement in $pp$ collisions 
in the same kinematical condition (\eg~the c.m.s. energy), $N_{pp}^{}$, the normalisation of such
a factor depends on an interpolation which introduces in additional systematical uncertainties. 
Because of this, it is interesting to look at an additional observable, 
the forward-to-backward ratio, defined as
\eqs{R_{\rm FB}(|y_{CM}|)\equiv \frac{dN_{pA}^{}(y_{CM})}{dN_{pA}^{}(-y_{CM})}=\frac{R_{p{\rm Pb}}(y_{CM})}{R_{p{\rm Pb}}(-y_{CM})}
\label{eq:RFB},}
in given rapidity and/or $P_T$ bins. Since
the yield in $pp$ is symmetric in $y_{CM}$, it drops out of the double ratio in the l.h.s. of \ce{eq:RFB}.

\vskip 0.25cm
In \cf{fig:fig1}, we present our results on open beauty for the nuclear modification factor $R_{pA}$ and 
the forward-to-backward ratio $R_{\rm FB}$ for $p$Pb collisions at 5 TeV.
These values can be compared to the measurements by the LHCb collaboration.
LHCb has reported~\cite{LHCb-CONF-2013-008} differential cross-sections for $J/\psi$ from $b$ 
as a function of $y$, both for $pA$ and $Ap$ collisions. 
>From them, we have extracted the corresponding forward-to-backward ratio, to be compared with our results. 
As can be observed in \cf{fig:fig1}, the measured values of $R_{p \rm Pb}$ seem to 
disfavour the presence of antishadowing. Indeed, the nDSg fit, which does not  
exhibit an antishadowing excess, matches the data points better.

However, we would like to  stress that $R_{p{\rm Pb}}$ is sensitive to the interpolated $pp$ cross section. 
To avoid this uncertainty, one may only want to focus on the comparison with the data for $R_{\rm FB}$. In such
a case, one does not see any tension with EPS09. 

\section{Conclusions}
We have presented our results for the open beauty nuclear modification factor in \pPb\ collisions
and its forward-to-backward ratio, $R_{\rm FB}$, at $\sqrt{s_{NN}}=5\mathrm{~TeV}$
as functions of $y$. Our results 
directly follow from the effects encoded in the nPDF which we have used
and can be compared to the LHCb data.

As for now, the precision of the measurement is limited by the collected statistics during the one month
2013 $p$Pb run, and by the absence of a $pp$ reference. Nevertheless, a comparison at the level of 
$R_{pPb}$ hints at an absence of gluon antishadowing, whereas a sole comparison at the level of 
$R_{\rm FB}$ does not hint at any disagreement. 

\section*{Acknowledgments}
We would like to thank R. Arnaldi, K. Eskola, C. Hadjidakis, F. Jing, G. Martinez, N. Matagne, F. Olness, H. Paukkunen, C. Salgado, I. Schienbein, R. Vogt, Z. Yang, and P. Zurita for stimulating and useful discussions. This work is supported in part by the Sapore Gravis Networking of the EU I3 Hadron Physics 3 program, by Ministerio de Economia y Competitividad of Spain (FPA2011-22776), by the French CNRS, grant PICS-06149 Torino-IPNO and by FEDER.


\end{document}